\documentclass[aps,prl,twocolumn]{revtex4}

\usepackage{graphicx}
\usepackage{dcolumn}
\usepackage{bm}
\usepackage{hyperref}
\usepackage{psfrag,bm,amsmath,amsfonts,amssymb,color}

\newcommand{\op}[1]{\hat{{\rm #1}}}

\newcommand{\sz}{\op{S}_z}
\newcommand{\spp}{\op{S}_+}
\newcommand{\smm}{\op{S}_-}

\newcommand{\IP}{\op{I}_+}
\newcommand{\IM}{\op{I}_-}
\newcommand{\Ix}{\op{I}_x}

\newcommand{\Piu}{\rm{P}_i^{\uparrow}}
\newcommand{\Pju}{\rm{P}_j^{\uparrow}}

\begin{document}


\title{Hyperfine interaction dominated dynamics of nuclear spins in self-assembled QDs}
\author{Christian Latta}\email{clatta@phys.ethz.ch}
\author{Ajit Srivastava}
\author{Atac Imamo\u{g}lu}\email{imamoglu@phys.ethz.ch}
\affiliation{Institute of Quantum Electronics, ETH-Z\"urich, CH-8093 Z\"urich, Switzerland.} 
\date{\today}

\begin{abstract}
We measure the dynamics of nuclear spins in a self-assembled quantum
dot at a magnetic field of 5 Tesla and identify two distinct
mechanisms responsible for the decay of the Overhauser field. 
We attribute a temperature-independent decay which lasts $\sim 100$ 
seconds to intra-dot  diffusion induced by hyperfine-mediated 
indirect nuclear spin interaction. In addition, we observe a 
gate-voltage and temperature dependent decay stemming from
co-tunneling mediated nuclear spin flip processes. By adjusting the
gate-voltage and lowering the electron temperature to $\sim 200$
milliKelvin, we prolong the corresponding decay time to $\sim 30$
hours. Our measurements indicate possibilities for exploring quantum 
dynamics of the central spin model using a single self-assembled quantum dot.
\end{abstract}

\maketitle


Hyperfine interaction between a single quantum dot (QD) electron and
the nuclear spin ensemble defined by the nano-scale confinement
provides a realization of the central spin problem~\cite{Prokofev2000,Gaudin1976,Al-Hassanieh2006,Dobrovitski2003,Coish2004,Chen2007}. This
model has attracted considerable attention recently since the
correlations between the confined electron and the nuclear spin
ensemble induced by hyperfine coupling constitute the principal
decoherence mechanism for the electron spin~\cite{Khaetskii2002}. It has been recognized
in this context that an enhancement of electron coherence time could
be achieved by preparing nuclear spins in eigenstates of the
Overhauser (OH) field operator~\cite{Reilly2008,Issler2010}: for this approach to be effective, it is
essential to understand and characterize the dynamics of prepared
(polarized) nuclear spin states.

In this Letter, we present measurements of nuclear spin
dynamics in a single electron charged self-assembled QD. In contrast to prior
work in self-assembled and gate-confined 
QDs~\cite{Maletinsky2007a,Latta2009,Chekhovich2010, Reilly2010}, we
probe nuclear spin dynamics when both the exchange coupling to a
Fermionic reservoir (FR) and the dipolar interaction between nuclear
spins are vanishingly small. Our observations reveal a spatially
limited, \textit{temperature-independent}, nuclear spin diffusion originating from electron
mediated nuclear spin interactions in addition to co-tunneling mediated, \textit{temperature dependent}, decay of the OH field approaching $10^5$~s.
Remarkably, the diffusion induced reduction in the OH field
taking place on $\sim$100 s timescale can be strongly
suppressed by repeating the preparation cycle consisting of
polarization (pump) and free-evolution (wait).

The QDs in this sample are separated by a 35
nm GaAs tunnel barrier from a doped n++-GaAs layer.
A bias voltage applied between a top semi-transparent Ti/Au
Schottky gate and back contacts allows to control the charging
state of the QD and the relative alignment of its electronic levels
with the Fermi energy of the FR~\cite{Warburton2000}. High-resolution 
resonance-scattering spectroscopy~\cite{Hogele2004} was performed on a
single QD in a fiber-based confocal microscope incorporated in a
dilution refrigerator~\cite{Latta2011}. The electron temperature was varied between
200 mK and 4 K while the applied magnetic field in the Faraday
geometry was kept constant at 5 Tesla. We adopted a modulation-free measurement
to keep the energy of the electron fixed during its interaction with
the nuclear spins.

\begin{figure}[ttt]
\begin{center}
   \includegraphics[scale=1]{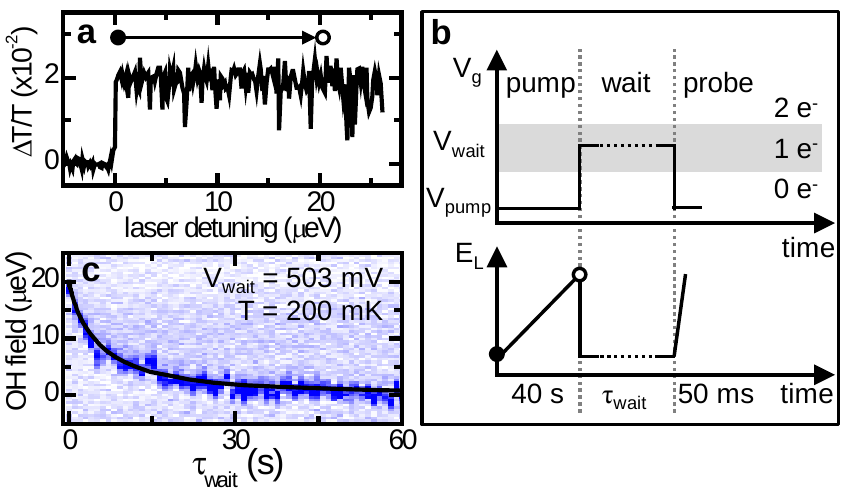}
    \caption{a) Build up of nuclear spin polarization by resonant optical dragging of the neutral exciton at 5T. b) Schematic of ``pump-probe" technique to measure the OH field decay: (1) A
    nuclear spin polarization is built up at a voltage $V_{pump}$ by slowly scanning the laser energy (full circle to open circle and back). (2) Gate voltage is set to $V_{wait}$
    for a time $\tau_{wait}$ keeping the laser energy fixed. (3)  Gate voltage is set to $V_{pump}$ again, followed
    by a fast (50 ms) laser scan to measure the OH field. c) A typical measurement of the OH
    field decay at 200 mK in the presence of a resident electron in the strong co-tunneling regime. The solid black line is calculated using the model (see text).\label{figure:NuclearSpinDiffusion:figure1}}
\end{center}
\end{figure}

We used a ``pump-probe" technique to investigate the OH
field dynamics. In the first step, the QD nuclear spins were
polarized by slowly scanning a single-mode laser across the blue
detuned Zeeman resonance of the neutral exciton ($X^0$): as was shown
in Ref.~\cite{Latta2009}, the magnitude of the OH field
obtained in such a ``dragging" experiment is given precisely by the
detuning of the applied laser field from the bare resonance. A typical dragging process is shown in
FIG.~\ref{figure:NuclearSpinDiffusion:figure1}a. After generating an OH field of
$\sim$20 $\mu$eV with $\sim$40 seconds of dragging, the gate 
voltage is abruptly changed (in $<1$ msec) to a
value $V_{wait}$ that results in the injection of an
electron into the QD from the FR (FIG.~\ref{figure:NuclearSpinDiffusion:figure1}b). 
Due to the large trion ($X^-$) energy-shift of $\sim$5~meV, the 
laser field is far off-resonance during the waiting time $\tau_{wait}$ in which the
coupled electron-nuclear system evolves freely. As a last step,
the magnitude of the remaining OH field is measured after
removing the electron from the QD and rapidly scanning the laser 
across the transition in $\sim$50 ms. This probing is fast 
enough not to cause any appreciable dynamic
nuclear spin polarization and simply reveals the resonance energy at
the time of the measurement.  We also confirmed
that no appreciable change in OH field takes place
during the time needed to switch the gate voltage between $V_{pump}$
and $V_{wait}$. The ``pump-wait-probe" sequence is then
repeated for different $\tau_{wait}$. A typical OH decay curve
obtained using this procedure is shown in FIG.~\ref{figure:NuclearSpinDiffusion:figure1}c.

\begin{figure}[ttt]
\begin{center}
   \includegraphics[scale=1]{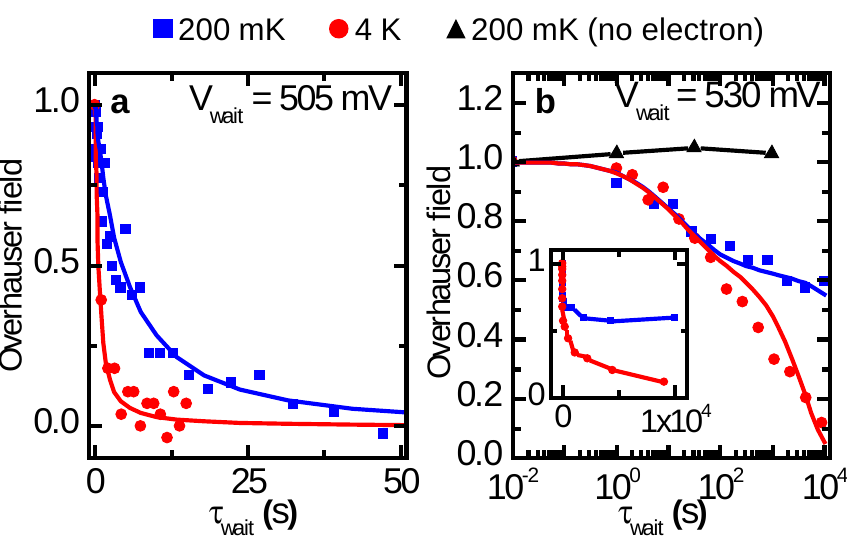}
    \caption{a) Decay of the OH field in a region of strong
    co-tunneling for $V_{wait}=505$~mV and different temperatures.
    b) Decay of the OH field with negligible co-tunneling for the case of 
    no resident electron ($\blacktriangle$),
    a resident electron at 200 mK ($\blacksquare$) and 4 K
    (\textbullet). The inset shows the same data in a
    linear-linear plot.\label{figure:NuclearSpinDiffusion:figure2}}
\end{center}
\end{figure}

In perfect agreement with earlier measurements~\cite{Maletinsky2009,Latta2009}, we found no
measurable decay of the OH field for an empty QD up to 1000 s. 
This result re-confirms that the OH field in
self-assembled QDs is stationary in the absence of a
confined electron (FIG.~\ref{figure:NuclearSpinDiffusion:figure2}b black triangles). 
A possible explanation for this observation is the presence of large and inhomogeneous
quadrupolar shifts within the QD that renders dipolar-interaction
mediated nuclear spin diffusion largely ineffective. Conversely,
the non-trivial OH field dynamics in the presence of an
electron that we discuss below demonstrates that the QD
electron-nuclei system is a near-perfect realization of the central
spin problem where the (nuclear) spins only interact with the 
central (electron) spin~\cite{Chen2007}.

When we choose $V_{wait}$ such that the single-electron charged
QD is in the co-tunneling regime\cite{Atature2006} and the exchange coupling to
the FR is strongest, we find that the OH field exhibits a fast decay~\cite{Maletinsky2007a,Latta2009} on the order of a few seconds. The observed decay is clearly temperature dependent (see FIG.~\ref{figure:NuclearSpinDiffusion:figure2}a).

In contrast, the OH field dynamics for $V_{wait}$ that corresponds 
to negligible electron co-tunneling shows decay on two distinct timescales (FIG.~\ref{figure:NuclearSpinDiffusion:figure2}). The initial decay now takes
place on a timescale of $\sim$100 s and is temperature independent.
Within this time, only a fraction of the OH field decays; for a
single polarization cycle the decaying fraction is approximately
$50\%$ (FIG.~\ref{figure:NuclearSpinDiffusion:figure2}b inset). 
This initial decay is followed by a
temperature-dependent slower decay which varies from $3500$ s at
4 K (FIG.~\ref{figure:NuclearSpinDiffusion:figure2}b red dots) to $10^5$~s at 200 mK
\footnote{Extrapolated from data taken up to $10^4$~s.} 
(FIG.~\ref{figure:NuclearSpinDiffusion:figure2}b blue squares).

\begin{figure}[ttt]
\begin{center}
   \includegraphics[scale=1]{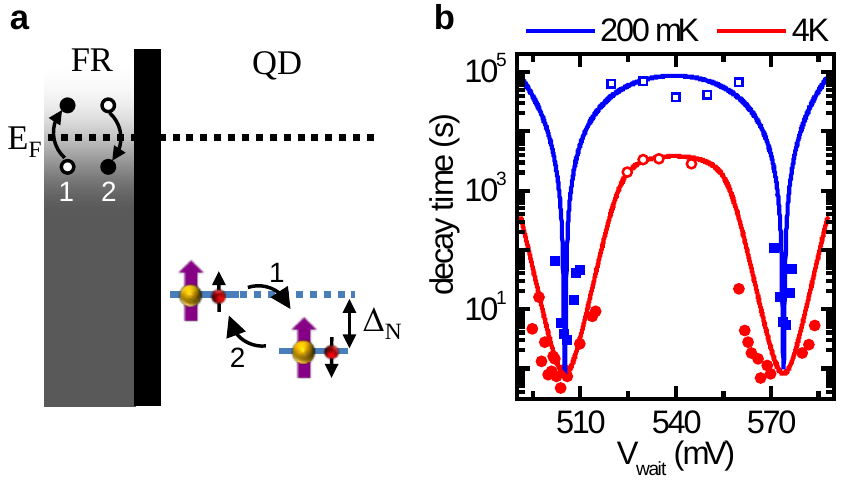}
    \caption{ a) Schematic of the temperature dependent
     decay process. A nuclear spin can be flipped without
      flipping the electron spin via non-collinear
      hyperfine interaction (see text). The energy difference
      $\Delta_N \ll T$ is provided (1) (absorbed (2)) via an
      particle-hole annihilation (excitation) in the FR.
      b)Extracted decay times of the OH field mediated by
      co-tunneling as a function of gate voltage.
      The solid lines are the calculated OH decay times (see text).
      \label{figure:NuclearSpinDiffusion:figure3}}
\end{center}
\end{figure}

The presence of two different OH field decay time scales
with different temperature dependence points to two independent
electron mediated mechanisms. To get further insight, we measured
the gate voltage dependence of the temperature dependent decay rate
across the single-electron charging plateau: in 
FIG.~\ref{figure:NuclearSpinDiffusion:figure3}, squares (circles) 
denote the experimentally measured values of the temperature dependent decay
time at $T=200$ mK ($T=4$ K). The values are extracted by fitting 
an exponential to this decay; the full (open) squares or circles 
indicate that the measured rate is the faster (slower) component 
of the OH field decay. The solid blue (red) curves show the gate voltage
dependence of the co-tunneling time at $T=200$~mK ($T=4$~K) scaled
by a (common) constant factor. The measured temperature dependent
decay rates follow the gate voltage and temperature dependence of the
co-tunneling rate across the charging plateau. In fact, we use
the expected linear temperature dependence of the
depicted co-tunneling rate and the good agreement with the
experimentally measured decay times to determine our electron
temperature to be $T \simeq 200$~mK\cite{Latta2011}~\footnote{The ratio 
of the fast decay rates at the
plateau edges for $T=4$~K and $T=200$~mK is $8$; this
discrepancy could arise due to gate voltage fluctuations that act as an effective
finite temperature when $V_{wait}$ is in the co-tunneling region.}.

\begin{figure}[ttt]
\begin{center}
   \includegraphics[scale=1]{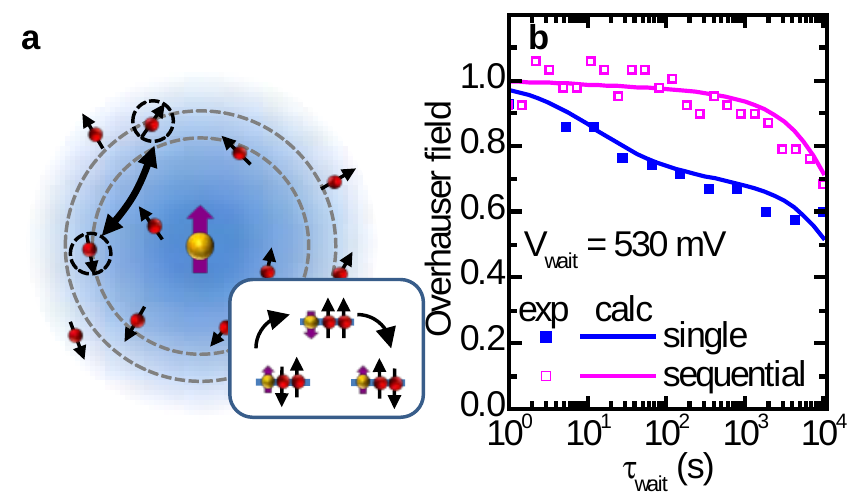}
    \caption{a) Schematic of the electron mediated nuclear
    spin diffusion: due to the inhomogeneous Knight shifts and
    quadrupolar fields, only nuclear spins with a small energy
    difference can interact which is depicted by the dashed
    circles. Two nuclear spins can flip without flipping 
    the electron spin. b) Demonstration of the spatially
    limited nuclear spin diffusion: By sequential
    polarization of the nuclear spins, the polarization can be
    saturated, suppressing further nuclear spin diffusion (see text for details).		 \label{figure:NuclearSpinDiffusion:figure4}}
\end{center}
\end{figure}

To investigate the temperature-independent initial decay of
the OH field, we polarized the QD nuclear spins successively in four
steps with a waiting time of 200~s in the presence of an electron
between each step. The magnitude of the
OH field at the end of each polarization cycle was kept the same. As shown in
FIG.~\ref{figure:NuclearSpinDiffusion:figure4}b, with 
successive polarization we find that the initial
decay of the OH field is practically eliminated. This
observation strongly suggests that the temperature independent component of decay
arises due to ``spatially limited diffusion" of nuclear spin
polarization. Indeed, with such a polarization scheme, the nuclear
spin polarization could diffuse within the QD in each dwell time
leading to an overall increase in the polarization. As a result, to
reach the same magnitude of the OH field in later steps,
progressively smaller nuclear polarization is required during dragging. 
As the diffusion process just redistributes the
excess polarization created during each step, one expects to see
smaller decay of the OH field with each
step, eventually leading to a complete suppression of the diffusion
induced decay of the OH field.

Our measurements thus indicate the presence of two qualitatively 
different mechanisms determining nuclear spin dynamics: temperature 
dependent decay and temperature independent diffusion of nuclear 
spin polarization, both mediated by the electron and leading to 
a decay of the OH field. To explain our findings, we use a
 model which includes Fermi-contact
hyperfine interaction ($\op{H}_{\rm{hyp}}$) describing the coupling of
the electron spin $\op{S}$ to $\sim 10^5$ nuclear
spins $\op{I}^i$ of the QD host material, exchange interaction
between the QD electron and the electrons in the FR
($\op{H}_{\rm{exch}}$), and an effective non-collinear dipolar hyperfine
interaction $\op{H}_{\rm{dip}}$. The total Hamiltonian can then be
written as $\op{H} = \op{H}_0 + \op{H}_{\rm{hyp}} + \op{H}_{\rm{exch}} + \op{H}_{\rm{dip}}$
where, $\op{H}_0 = \Delta\op{S}_z + \sum_i{\Delta_N^i \op{I}^i_z}$
is the Zeeman-Hamiltonian with $\Delta = g_e\mu_B B\approx 180\mu
eV$ and $\Delta_N^i$ denoting the electronic and nuclear Zeeman
energies, respectively. Here we have incorporated the inhomogeneous quadrupolar
interaction induced energy shifts of each nucleus in $\Delta_N^i$;
since typical quadrupolar fields ($0.3$ Tesla) are much smaller than
the external field ($5$ Tesla), we have $\Delta_N^i \simeq \Delta_N
= g_N \mu_N B \approx 0.1\mu eV$. $\op{H}_{\rm{exch}} =
\sum_{k,k'}{J_{k,k'}\op{s}_{k,k'}\cdot\op{S}}$ describes the
exchange coupling between the QD electron and the FR
with $\op{s}_{k,k'}$ denoting the spin operator of the FR at the
position of the QD and $J_{k,k'}$ denoting the exchange interaction
strength. This interaction leads to an incoherent electron spin flip
rate $\kappa$ which in our QD is $\sim 10^{-7}\mu\rm{eV}$ at 4 K in the center
of the charging plateau where the co-tunneling is smallest.

The Fermi-contact hyperfine interaction is given by $\op{H}_{\rm{hyp}} = \sum_i{A_i(\op{I}^i_z \op{S}_z
+\frac{1}{2}(\IP^i\smm + \spp\IM^i))}$, where $A_i\propto|\psi(\vec{r}_i)|^2\approx
10^{-2}\mu eV$ is the hyperfine constant of the $i$-th nucleus with
$\psi(\vec{r}_i)$ denoting the QD electron wave-function.
The first term in $\op{H}_{\rm{hyp}}$ is the OH (Knight) field
seen by the electron (nuclei).
For large magnetic fields used in our experiments,
the flip-flop terms in $\op{H}_{\rm{hyp}}$ are ineffective due to the large
difference in the electron and nuclear Zeeman energies.
Eliminating these terms in Eq.~(3) using a Schrieffer-Wolff transformation
\cite{Klauser2008}, we obtain new terms describing
electron mediated spin flip of two spatially separated nuclear spins:
\begin{equation}
    \op{H}_{\rm{ind}} = \sum_{i \neq j}{\frac{2A_iA_j}{\Delta-\Delta_N}\IM^i\IP^j\sz}  \;\; .
\end{equation}
This indirect, coherent long-range interaction leads to a diffusion of nuclear
spin polarization within the region where the electron wave-function
is non-vanishing (FIG.~\ref{figure:NuclearSpinDiffusion:figure4}a). 
Although the total magnitude of QD nuclear spin
polarization does not decrease due to this diffusion process, the
OH field seen by the electron decays partially due to a 
redistribution of the nuclear spin polarization within 
the QD. We attribute the temperature independent decay 
of the OH field to such an electron mediated diffusion.

The last term in the Hamiltonian describes an effective
non-collinear dipolar hyperfine interaction between the electron
and the $i$- th nucleus with coupling constant $B_i$:
\begin{equation}
\op{H}_{\rm{dip}} = \sum_i{B_i\Ix^i\sz} \;\; .
\end{equation}
Such terms could appear due to small but non-zero dipolar hyperfine
interaction between the QD electron and nuclei. Alternatively,
they could be induced by quadrupolar axes of nuclear spins that are
non-parallel to the external field~\cite{Huang2010}. These terms
induce nuclear spin flips that lead to a decay of the nuclear spin
polarization. In fact, the temperature-dependent decay of the OH
field can be explained by a second order process originating from $\op{H}_{\rm{dip}}$.
The energy conservation in this irreversible nuclear spin flip
process is ensured by the coupling of the QD electron to the FR (FIG.~\ref{figure:NuclearSpinDiffusion:figure3}a); the
corresponding OH field decay rate is then
$(B_i/\Delta_N)^2\kappa$. The explains the temperature and gate voltage
dependence of the decay shown in FIG.~\ref{figure:NuclearSpinDiffusion:figure3}.
Since $\Delta_N \ll T$, we expect this
rate to be linearly proportional to the electron temperature $T$. 
We rule out any contribution of co-tunneling assisted 
direct hyperfine flip-flop processes, since the 
corresponding rate can be shown to be four-orders-of-magnitude 
slower than the rates that we measure in our experiments.

We model the nuclear spin dynamics using semi-classical rate 
equations, taking into account
the diffusion and decay processes arising from $\op{H}_{\rm{ind}}$ and
$\op{H}_{\rm{dip}}$, respectively. For simplicity, we
assume a two-dimensional QD with N = $10^4$ spin 1/2 nuclei. 
The rate equations describing the change in 
time of the probability $\Piu(t)$ that the
i-th nucleus is in the $|\uparrow\rangle$ state become:
\begin{eqnarray}
\frac{d\Piu(t)}{dt} &=& \left(\frac{B_i}{\Delta_N}\right)^2\kappa(1-2\Piu(t)) \nonumber\\
&+& \sum_j{\left(\frac{2A_iA_j}{\Delta}\right)^2\rho_{ij}
        (\Pju(t)-\Piu(t))}.
\label{MasterEqn}
\end{eqnarray}
The first term on the RHS of Eq.~\ref{MasterEqn} represents 
the temperature dependent decay while the
second term represents the decay induced by the spatially limited
diffusion. To obtain Eq.~\ref{MasterEqn}, we assume that the
coherent coupling of two distant nuclear spins with similar energies
via $\op{H}_{\rm{ind}}$ is interrupted by a pure dephasing process
with rate $\gamma_{deph}$. The Lorentzian factor $\rho_{ij} =
\gamma_{deph}/(\delta_{ij}^2 + \gamma_{deph}^2)$ describes the
effective density of states for the flip-flop process between two
(distant) nuclear spins with energy difference $\delta_{ij} =
\Delta_N^i + A_i - \Delta_N^j + A_j$.

A possible source of $\gamma_{deph}$ is the intrinsic gate
voltage fluctuations in our experimental setup; such
fluctuations would influence the electron wave-function giving rise
to an effective broadening of the Knight field experienced by the
nuclei \footnote{Charge fluctuations near the QD can also
induce fluctuating electric field gradients and cause a broadening
of nuclear spin energies; this process will also contribute to
$\gamma_{deph}$.}. As the bandwidth of this noise is limited by
the bandwidth of the gate in our sample ($\sim$50~kHz), these 
fluctuations should not affect the decay process which is 
accompanied by a nuclear spin flip and requires an energy 
exchange of $\sim \Delta_N$ $\approx$ 0.1 $\mu eV$. Eq.~\ref{MasterEqn} 
can be solved for a given initial distribution of nuclear 
polarization which we assume is proportional to $\psi(\vec{r})$; 
with the knowledge of $\Piu(t)$ for
all nuclear spins, one can easily get the OH field as OH(t)
= $\sum_{i}A_{i}(\Piu(t)-0.5)$. For the calculations, we 
used $\Delta$=$174~\mu$eV, $\Delta_{\rm{N}}$=$0.1\mu$eV, 
$\sum{A_i}$=$174~\mu$eV (5 T) and $B_i\sim 10^{-2}A_i$. We first fix the parameters for the case 
of smallest co-tunneling at 4 K ($\kappa=10^{-7}~\mu$eV) and then use calculated 
$\kappa(V_{wait}, T)$ to obtain OH(t) for different temperatures and gate voltages. We obtain a value of $\sim$ 2 kHz for $\gamma_{deph}$  which is well below the bandwidth of the gate. The results of the calculations plotted in FIGS.~\ref{figure:NuclearSpinDiffusion:figure1}-~\ref{figure:NuclearSpinDiffusion:figure4} with solid lines, show good agreement with the experiment.

Our results demonstrate that the nuclear spin dynamics is solely
determined by the coupling of each nucleus to the central electron
spin. At ultra-low temperatures, the OH field decay for a QD well
isolated from an electron reservoir is predominantly due to
intra-dot diffusion. By saturating the diffusion process using
multiple polarization cycles and reducing the extrinsic (gate
voltage) fluctuations that enhance the diffusion rate, it should be
possible to prolong the spin-echo $T_2$ time of the electron
spin~\cite{Taylor2007}. In addition, elimination of the dephasing of
indirect interaction would open up the possibility for observation
of coherent quantum dynamics of the nuclear spins upon an abrupt
turn-on of the Fermi-contact hyperfine interaction.

\end{document}